\documentclass[journal]{IEEEtran}
\usepackage[dvipsnames]{xcolor}
\usepackage{colortbl}
\usepackage{cite}
\usepackage{amsmath,amssymb,amsfonts}
\usepackage{graphicx}
\usepackage{textcomp}
\usepackage{graphicx}
\usepackage{graphicx}
\usepackage{import}
\usepackage{amsmath,graphicx}
\usepackage{soul}
\usepackage{amssymb}

\usepackage{times}

\usepackage{tabularx,booktabs}
\usepackage{arydshln}
\usepackage{tabu}
\usepackage{adjustbox}
\usepackage{enumerate,multicol}
\usepackage{multirow}
\usepackage{multicol}
\usepackage{enumitem}

\usepackage{amsmath,array,graphicx}
\usepackage{gensymb}
\usepackage{float}
\usepackage{xspace}
\usepackage{amssymb}
\usepackage[algo2e,ruled]{algorithm2e}
\usepackage{algorithm}
\usepackage{pifont}
\usepackage{subcaption}
\usepackage{tabularx}
\usepackage{hyperref}
\usepackage{amsmath}
\usepackage{mathtools}
\usepackage{xfrac}
\hypersetup{
    colorlinks,
    linkcolor={red!50!black},
    citecolor={blue!50!black},
    urlcolor={black}
}

\usepackage{array}
\newcolumntype{P}[1]{>{\centering\arraybackslash}p{#1}}

\makeatletter
\newcommand\avsuminner[2]{%
  {\sbox0{$\m@th#1\sum$}%
   \vphantom{\usebox0}%
   \ooalign{%
     \hidewidth
     \smash{\vrule height\dimexpr\ht0+1pt\relax depth\dimexpr\dp0+1pt\relax}%
     \hidewidth\cr
     $\m@th#1\sum$\cr
   }%
  }%
}
\definecolor{shadecolor}{rgb}{0.9,0.9,0.9}
\usepackage{times}

\usepackage{soul}
\usepackage{url}
\usepackage[]{hyperref}
\usepackage[utf8]{inputenc}
\usepackage{amsmath}
\usepackage{booktabs}
\usepackage{flushend}
\DeclareMathOperator*{\argmax}{argmax}

\def\BibTeX{{\rm B\kern-.05em{\sc i\kern-.025em b}\kern-.08em
    T\kern-.1667em\lower.7ex\hbox{E}\kern-.125emX}}
\begin{document}

\title{Predicting Postoperative Intraocular Lens Dislocation in Cataract Surgery via Deep Learning}

\author{
\IEEEauthorblockN{
Negin Ghamsarian\IEEEauthorrefmark{1}, 
Doris Putzgruber-Adamitsch\IEEEauthorrefmark{3}, 
Stephanie Sarny\IEEEauthorrefmark{3}, 
Raphael Sznitman\IEEEauthorrefmark{1}, 
Klaus Schoeffmann\IEEEauthorrefmark{2}, and 
Yosuf El-Shabrawi\IEEEauthorrefmark{3}\IEEEauthorrefmark{4}
}

\IEEEauthorblockA{\IEEEauthorrefmark{1} ARTORG Center for Biomedical Engineering Research, University of Bern, Bern, Switzerland \\\{negin.ghamsarian, raphael.sznitman\}@unibe.ch}

\IEEEauthorblockA{\IEEEauthorrefmark{2} Department of Information Technology, Klagenfurt University, Klagenfurt, Austria\\ ks@itec.aau.at}

\IEEEauthorblockA{\IEEEauthorrefmark{3} Department of Ophthalmology, Klinikum Klagenfurt, Klagenfurt, Austria \\ \{doris.putzgruber-adamitsch,stephanie.sarny\}@kabeg.at}

\IEEEauthorblockA{\IEEEauthorrefmark{4} Department of Ophthalmology, Medical University of Graz, Graz, Austria \\ yosuf.elshabrawi@medunigraz.at}

\thanks{This work was funded by the Haag-Streit Foundation, Switzerland, and the FWF Austrian Science Fund under grant P 31486-N31.}
}




\maketitle

\begin{abstract}
A critical yet unpredictable complication following cataract surgery is intraocular lens dislocation. Postoperative stability is imperative, as even a tiny decentration of multifocal lenses or inadequate alignment of the torus in toric lenses due to postoperative rotation can lead to a significant drop in visual acuity. Investigating possible intraoperative indicators that can predict post-surgical instabilities of intraocular lenses can help prevent this complication. In this paper, we develop and evaluate the first fully-automatic framework for the computation of lens unfolding delay, rotation, and instability during surgery. Adopting \textcolor{black}{a combination of} three types of CNNs, namely recurrent, region-based, and pixel-based, the proposed framework is employed to assess the possibility of predicting post-operative lens dislocation during cataract surgery. This is achieved via performing a large-scale study on the statistical differences between the behavior of different brands of intraocular lenses and aligning the results with expert surgeons' hypotheses and observations about the lenses. We exploit a large-scale dataset of cataract surgery videos featuring four intraocular lens brands. Experimental results confirm the reliability of the proposed framework in evaluating the lens' statistics during the surgery. The Pearson correlation and t-test results reveal significant correlations between lens unfolding delay and lens rotation and significant differences between the intra-operative rotations stability of four groups of lenses. These results suggest that the proposed framework can help surgeons select the lenses based on the patient's eye conditions and predict post-surgical lens dislocation. 

\end{abstract}

\section{Introduction}
Cataract refers to the cloudiness of the eye's natural lens, usually due to aging, resulting in vision blur, dimness, distortion, double vision, and degraded color perception. Cataracts are the major cause of blindness worldwide~\cite{VB2020}. 
\textcolor{black}{Due to the aging population and longer life expectancies, the World Health Organization (WHO) predicts that the incidence of cataract-related blindness will rise to 40 million by 2025~\cite{ITBIOL}.}
This common disease can be remedied by replacing the natural lens with an artificial lens termed intraocular lens (IOL) during cataract surgery~\cite{OBCS,MOCS}. Cataract surgery is the most frequent eye surgery and one of the most frequently performed surgeries worldwide. 
\textcolor{black}{With the continuous introduction of technological advancements, cataract surgery techniques are constantly evolving. The progression has witnessed significant shifts, starting from intracapsular cataract extraction (ICCE) in the 1960s and 1970s, to extracapsular cataract extraction (ECCE) in the 1980s and 1990s, and currently, the widely adopted technique is sutureless small-incision phacoemulsification surgery with injectable intraocular lens (IOL) implantation\footnote{In this paper, the term "Cataract Surgery" refers to "Phacoemulsification Cataract Surgery."}. These advancements in surgical methods demonstrated tangible enhancements in visual outcomes and safety~\cite{ITBIOL,ECSS}.}
Although not exceeding $10\%$ with mostly transient effects, the intra-operative and post-operative complications in cataract surgery may lead to visual impairment and severe patient discontent~\cite{AppCat,RPhC}\textcolor{black}{, \cite{day2015royal,gross2004bag}}. Due to the prevalence of cataract surgery and its considerable impact on the patient's quality of life, predicting and avoiding its post-operative complications is of prime concern for the surgical community. 

Intraocular lens dislocation is a major post-operative complication following cataract surgery~\textcolor{black}{\cite{RFIOLD,subasi2019late,kristianslund2021late}}. During the procedure, the eye's natural lens is removed  
and a folded artificial lens (IOL) is inserted into the eye's capsular bag. The lens then unfolds and possibly rotates and dislocates until completely unfolded. Despite being aligned and centralized at the end of the surgery, in some cases, the IOL rotates or dislocates following the surgery. Even minor misalignments of the torus in toric IOLs and decentration and tilting of multifocal IOLs can lead to significant vision distortion and dissatisfied patients. Follow-up surgery is currently the only way to address this post-operative complication, \textcolor{black}{entailing} additional costs, surgical risks, and patient discomfort. There is an unmet clinical demand to identify intra-operative indicators to predict and avoid this post-operative complication during the surgery. 

It is argued that early rotation in toric IOLs during cataract surgery is the leading cause of post-operative misalignments~\cite{Oshika2020}. Since the unfolding delay differs between various IOL brands, it is hypothesized that there is a direct correlation between the lens' behavior during unfolding and its post-operative stability.  Besides, an incomplete unfolding of the IOLs at the end of surgery may lead to an inadequate pressure of the haptics against the capsular bag, thus resulting in post-operative rotation, decentration, or tilting of the IOL. In recent years, extensive research has been conducted to compare and predict the rotation stability of different IOLs~\cite{LEE20181325, SCHARTMULLER202072, INOUE20171424, Schartmuller186, Mayer-Xanthaki1510}. However, a reliable study \textcolor{black}{for} evaluating the behavior of IOL during its unfolding or other risk factors during the surgery requires large-scale comparisons, necessitating an automatic lens' feature extraction method from surgical microscope video feeds.

In this paper, we aim to investigate the possibility of automating the statistical analysis \textcolor{black}{for} different intraocular lens (IOL) behaviors during surgery to predict post-operative lens rotational stability. \textcolor{black}{The main contributions of this paper are as follows.}

\begin{enumerate}

    \item \textcolor{black}{We introduce the first deep-learning-based framework for automatic analysis and comparison of four brands of IOLs based on (i) lens unfolding delay, (ii) lens instability, and (iii) lens rotation during the surgery. To achieve this, three deep-learning-based architectures are employed to tackle different problems in surgical scene understanding: (a) a recurrent convolutional neural network for precise implantation phase detection, (b) a U-Net-based network for lens and pupil segmentation after the implantation phase, and (c) a region-based network for lens' hook detection.}
    \item \textcolor{black}{The proposed framework is evaluated using a large-scale dataset of cataract surgery videos.}

    \item \textcolor{black}{Using the proposed framework with trained models, a large-scale study is conducted based on the statistics of four groups of intraocular lenses.}  
    \item \textcolor{black}{The fully-automated statistical comparisons among these four brands of intraocular lenses for the first time suggest significant correlations between lens unfolding delay and rotation and significant differences among the rotation degrees of different lenses. 
    }
\end{enumerate}
\textcolor{black}{The efficacy of each stage in the proposed framework is evaluated using relevant metrics, including (I) precision, recall, f1-score, and accuracy for phase recognition, (II) Jaccard metric and dice coefficient for semantic segmentation, and (III) mean average precision for object detection and pose estimation.
The phase recognition network achieved $100\%$ accuracy in detecting the implantation phase, which is the starting time to compute lens unfolding delay and rotation. The segmentation network showed outstanding performance in lens and pupil segmentation (a dice coefficient equal to $92.62\%$ for lens segmentation and $97.98\%$ for pupil segmentation). Ultimately, the proposed lens orientation calculation method demonstrates a mean error as small as 3.707 degrees, confirming the detections' high reliability.
Our statistical evaluation results align with the surgeons' hypotheses regarding the correlation between lens behaviors during and after surgery. By demonstrating the possibility of predicting and subsequently reducing post-operative complications of intraocular lenses through lens behavior evaluation during surgery, our results provide evidence for potential improvements in patient outcomes.}

The rest of this paper is organized as follows. In Section \ref{sec: related work}, we position our work in the literature by reviewing state-of-the-art methods on artificial-intelligence-assisted analysis of cataract surgery videos. Section \ref{sec: Methodology} details our proposed framework for computing the IOL statistics during the surgery. We explain the experimental setup in Section \ref{sec: experimental setup} and present the experimental results in Section \ref{sec: experimental results}. Finally, Section \ref{sec: discussion} discusses the achievements of our work and concludes the paper.

\begin{figure*}[!tb]
    \centering
    \includegraphics[width=1\textwidth]{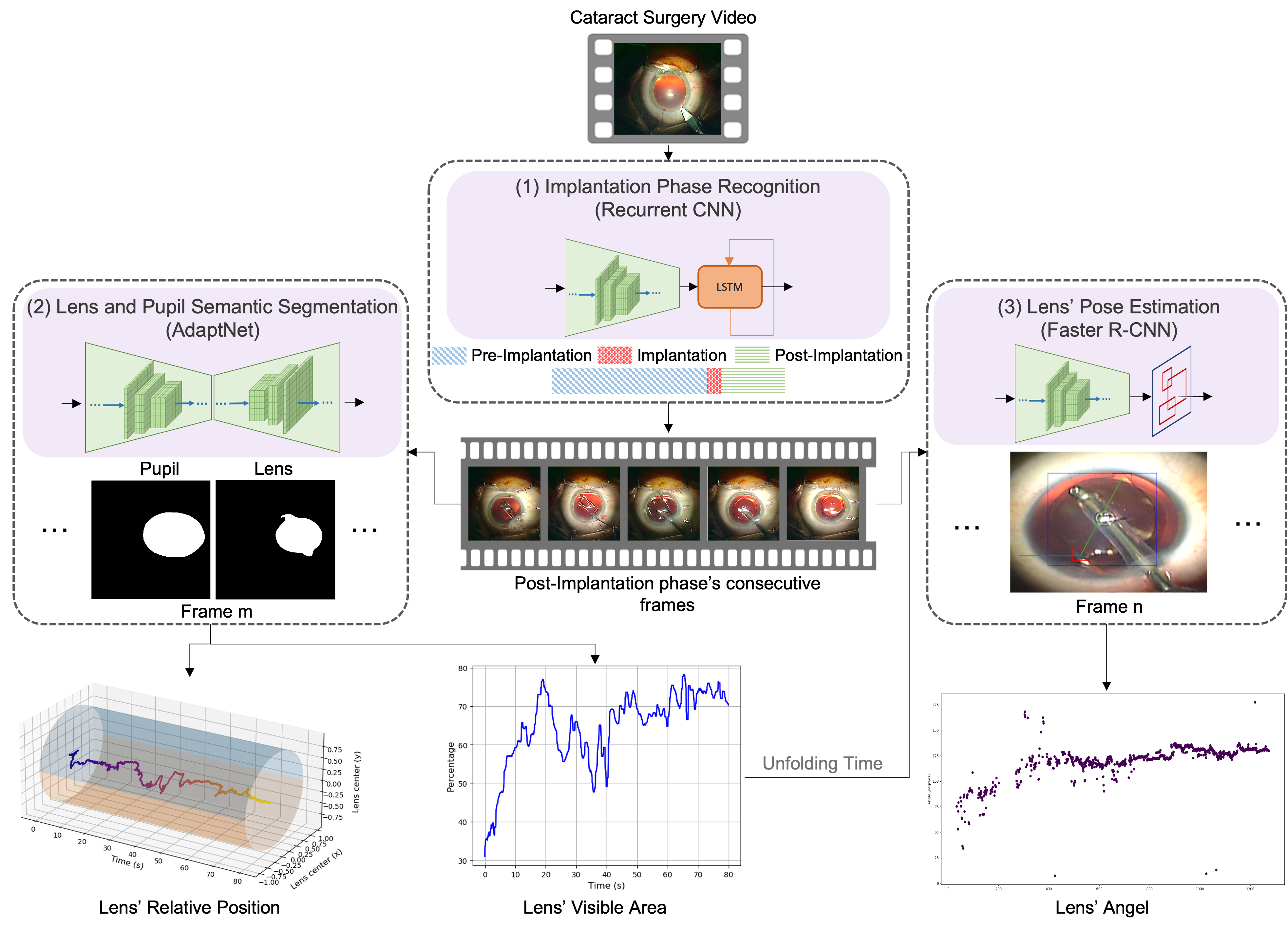}
    \caption{The overall framework of the proposed intraocular lens qualification method. In the first stage, a recurrent CNN detects the implantation phase to label the last frame of this phase as the starting point for intraocular lens statistics computation. Afterward, a semantic segmentation network outputs the masks of the intraocular lens and pupil to be used for lens unfolding delay and instability computation. Finally, a region-based CNN is employed to compute absolute lens rotations after full unfolding until the end of the video.}
    \label{fig:ID}
\end{figure*}

\section{Related Work}
\label{sec: related work}



\textcolor{black}{
The field of cataract surgery has witnessed the integration of artificial intelligence (AI) to address a spectrum of demands spanning pre-operative, intra-operative, and post-operative applications. Regarding pre-operative requisites, AI has been instrumental in supporting surgical diagnosis and decision-making, including cataract detection and grading, as evidenced by numerous studies~\cite{ADSNC, AICRC, EaAAI, AGNC}. Classic AI-based methods for the intra-operative and post-operative applications focused on instrument tracking~\cite{PhacoTracking}, surgical process modeling~\cite{SurgPM}, surgical training~\cite{RemSup,ViRe,ORPI,FHF}, robot-assisted surgery~\cite{RASCS,SAOGCR}, and surgical time prediction~\cite{PredSurg}. Furthermore, AI has demonstrated its effectiveness in predicting outcomes related to cataract surgery, notably in the calculation of intraocular lens power~\cite{AoaNIOL}.}

\begin{figure*}[!tb]
    \centering
    \includegraphics[width=0.97\textwidth]{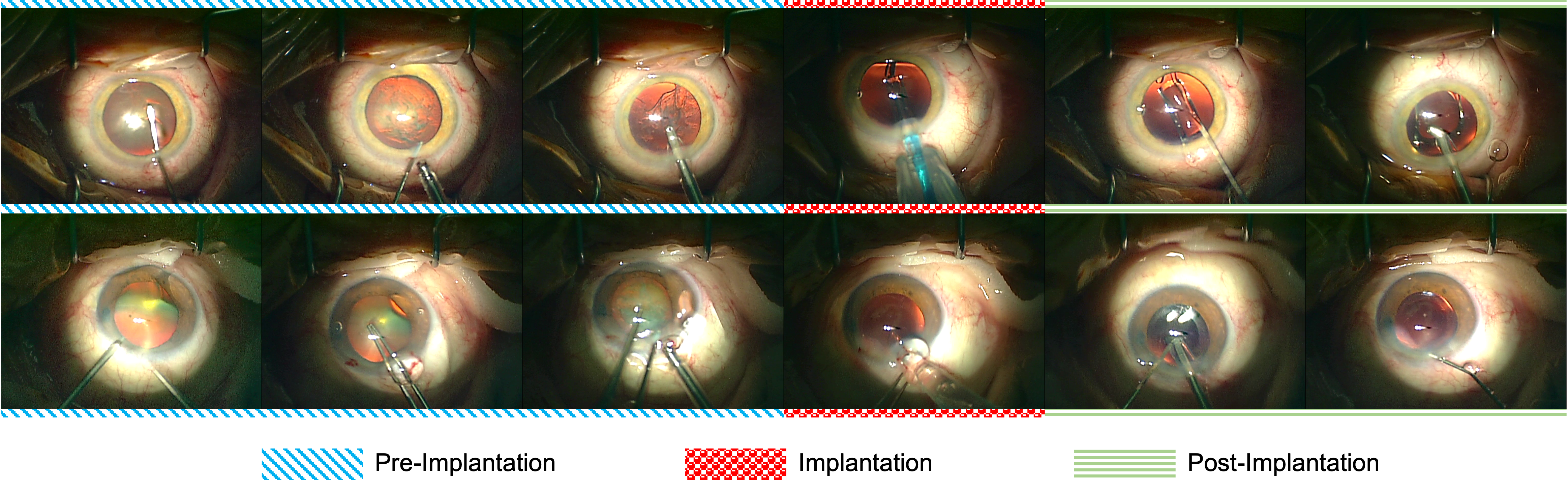}
    \caption{Sample frames from the pre-implantation, implantation, and post-implantation phases from two representative videos.}
    \label{fig: cataract surgery sample frames}
\end{figure*}

\textcolor{black}{In recent years, convolutional neural networks (CNN) have become the predominant driving engine in computerized surgical workflow analysis. Recent studies have showcased the capabilities of CNN-based frameworks in pre-operative cataract diagnosis \cite{tham2022detecting,zeboulon2022development,tripathi2022mtcd}, and cataract type and severity classification \cite{DeepLensNet}. 
The intra-operative deep-learning-based methods can be categorized into two primary areas (i) operation room planning and (ii) intra-operative surgical guidance. Operation room planning encompasses tasks like
predicting remaining surgery duration~\cite{CataNet,wang2023real} and surgical site confirmation \cite{yoo2020deep}. Real-time guidance in particular phases \cite{EoAIBIGT} and pupil reaction detection \cite{ADPR} are examples of the latter group of methods. 
Post-operative cataract surgery analysis methods primarily focus on surgical training and prognosis. Workflow analysis methods are integral in this context and include but are not limited to CNN architectures for phase classification~\cite{AAIP,wang2022intelligent,touma2022development}, joint phase segmentation-classification~\cite{RDCSV}, instrument tracking \cite{yeh2023phacotrainer}, and deblurring cataract surgery videos \cite{DCS}. Furthermore, several CNN-RNN-based frameworks have been proposed to perform relevance-based compression~\cite{RelComp}, and surgical training expedition\cite{RBE,whitten2022clinically}. Recent studies have also explored automated technical skill assessment in robotic surgeries \cite{ruzicki2023use}. 
Given the fundamental role of semantic segmentation in various surgical workflow analysis applications, significant efforts have been invested in improving semantic segmentation performance in cataract surgery \cite{DeepPyramid,ReCal-Net,LensID}. In recent years, some efforts have been made to enable surgical prognosis, such as post-surgical visual acuity prediction \cite{wei2021optical}. These developments collectively underscore the transformative role of CNNs in advancing the field of cataract surgery and its associated analysis techniques.}

\textcolor{black}{Regarding research about intraocular lenses, substantial attention has been dedicated to various pre-operative methods, including IOL power calculation~\cite{SupFo,UoIOLCF,MLICS}, IOL parameter verification \cite{kiuchi2022deep}, and IoL segmentation \cite{schwarzenbacher2022automatic}. 
However, when it comes to the prognosis of cataract surgery, the role of artificial intelligence remains relatively underexplored, with limited work on predicting posterior capsule opacification~\cite{PCO}.}

\textcolor{black}{This study aims to leverage the CNNs' power to predict lens dislocation as a significant postoperative complication in cataract surgery. 
Our primary objective is to develop a fully automatic framework capable of computing critical parameters such as lens unfolding delay, rotation, and instability from cataract surgery videos. By achieving this, we aim to facilitate an in-depth analysis that compares intra-operative statistics across various brands of IOLs. Moreover, this research serves as a foundational step toward predicting and preventing post-operative lens dislocation, a significant concern in cataract surgery.
}

\section{Methodology}
\label{sec: Methodology}

Figure 1 demonstrates the pipeline of the proposed method for automatic lens statistic computation during cataract surgery. The pipeline mainly consists of three modules: (1) implantation phase recognition, (2) lens and pupil semantic segmentation, and (3) lens' pose estimation. For the two first modules, we use our proposed neural network architectures~\cite{LensID}. In this section, we detail the functionality of each module in the proposed framework. We then explain the lens statistic computation and correlation analysis in \ref{sec: lens statistic computation} and \ref{sec: statistical analysis}, respectively. The pseudocode of our proposed framework for IOL evaluation is present in \textbf{Algorithm}~\ref{algorithm}.

\subsection{Implantation-Phase Recognition}

As the first step toward lens evaluation, we set the starting point for lens statistics computation to be the post-implantation phase, where the folded IOL is inserted inside the eye using a cartridge. Figure \ref{fig: cataract surgery sample frames} illustrates randomly sampled frames from pre-implantation, implantation, and post-implantation phases for two representative videos. We utilize a recurrent CNN with a many-to-many architecture to detect the implantation phase accurately. Recurrent convolutional neural networks can detect the label associated with a sequence of input frames considering the intertwined spatiotemporal features. Moreover, by exploiting features from the neighboring frames, recurrent CNNs can mitigate degraded frame quality typical in cataract surgery videos, such as harsh motion blur and defocus blur. We exploit a stochastic sampling strategy during training to avoid network overfitting to the speed and skill level of the surgeons and improve the network's generalization performance. 

More specifically, the network detects the associated phase label to each three-second clip as follows: (1) the three-second sequence with the rate of 25 frames per second is split into five subsequences, each containing 15 consecutive frames; (2) a frame is randomly sampled as the keyframe from each subsequence; (3) the five sampled frames are fed to the network, and for every frame, the network outputs \textcolor{black}{the} probability of belonging to the implantation phase and (4) the output probabilities are averaged to obtain the predicted label for the three-second input sequence. During inference, we perform uniform sampling to provide better diversity in the input frames. Following the detection of the labels for all consecutive three-second clips, we have the time-slot range of the pre-implantation, implantation, and post-implantation phases. We use the post-implantation phase for computing the IOL statistics.

\begin{figure}[!tb]
    \centering
    \includegraphics[width=1\columnwidth]{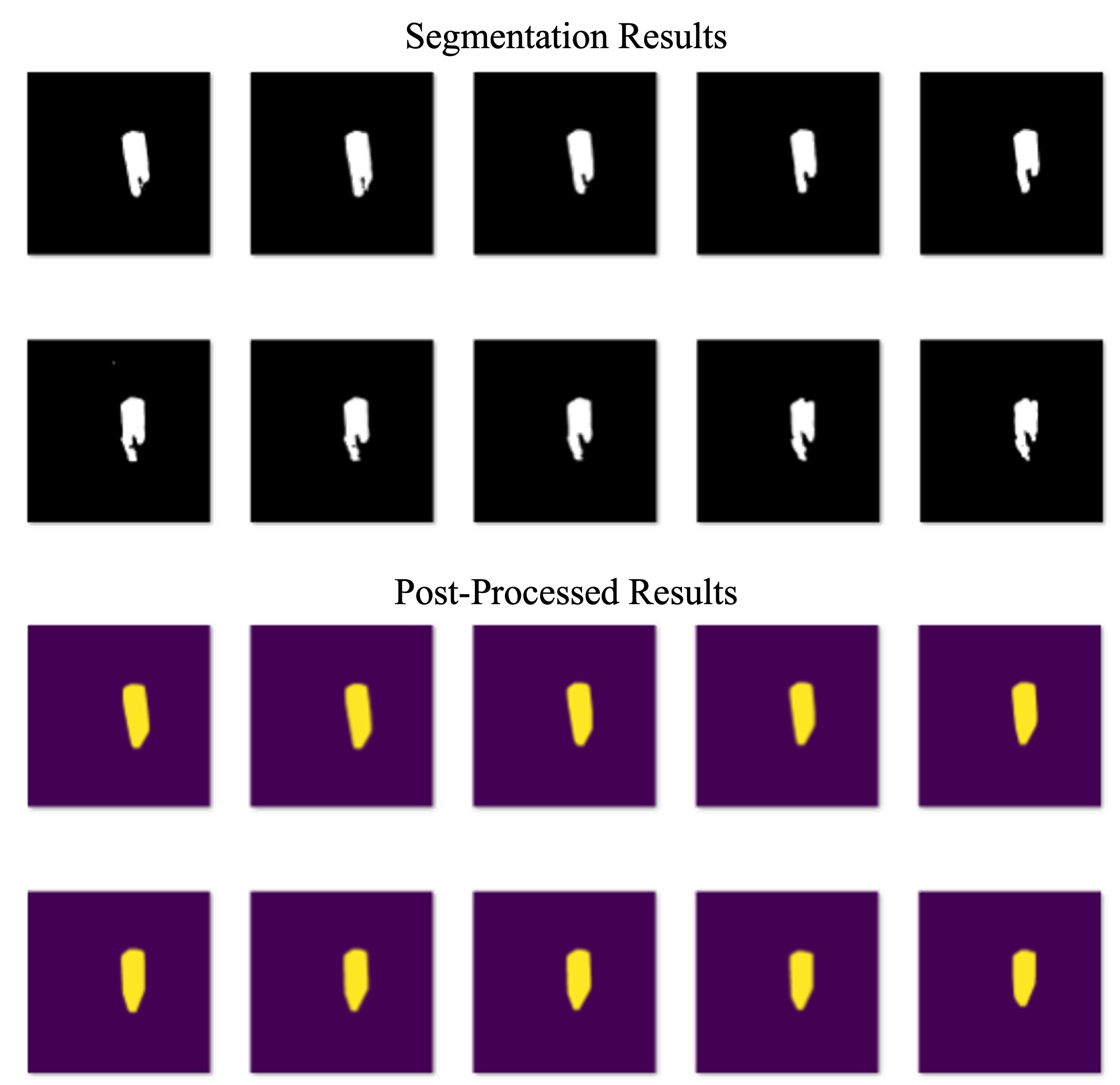}
    \caption{The IOL segmentation results for ten consecutive frames containing dents in the regions of instrument occlusion, and their corresponding refined versions where occluded segments are recovered.}
    \label{fig:Segmentation refinement}
\end{figure}
\subsection{Lens and Pupil Semantic Segmentation}
To compute the lens unfolding time and instability, we track changes in the size of the IOL over time. Accordingly, we require a semantic segmentation mask of the pupil and IOL, for which we use the AdaptNet architecture~\cite{LensID}. 
This network takes advantage of some novel modules to induce shape and scale awareness of the network. These modules can effectively deal with various difficulties in segmenting the IOL due to its transparency, unpredictable formation during unfolding, occlusion by the instruments, defocus blur, and motion blur.

\begin{figure*}[!tb]
    \centering
    \includegraphics[width=0.95\textwidth]{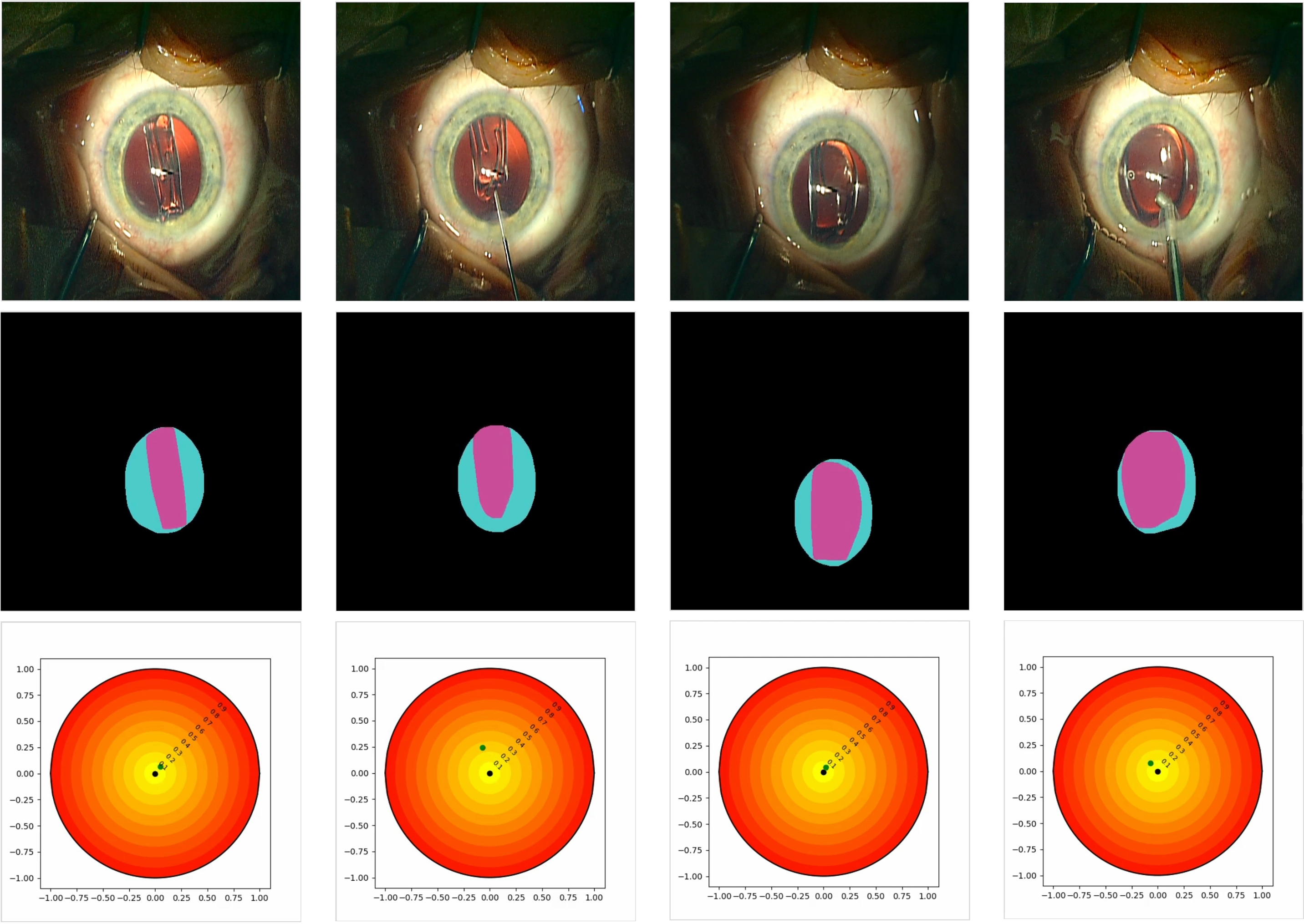}
    \caption{Visualization of lens unfolding, rotation, and instability for a representative video. The frames from left to right show the IOL unfolding procedure. Top: sample frames from the post-implantation phase; Middle: the corresponding refined IOL and pupil masks;  Bottom: the lens center relative to the pupil center corresponding to each frame.}
    \label{fig:lens instability}
\end{figure*}

The lens and pupil segmentation results are then post-processed. Since there are several phases after implantation, the lens and pupil are usually occluded with the instruments, and the segmentation networks cannot detect the occluded regions. However, these regions should be included in the area of the lens and pupil. We adopt the domain-specific knowledge related to these objects to retrieve the occluded regions in their semantic segmentation masks. Specifically, since the IOL and pupil are often convex objects, we draw a convex polygon around each detected object to retrieve the occluded regions. Figure \ref{fig:Segmentation refinement} compares the segmentation results and refined masks for some consecutive frames.
After post-processing, the visible area of the IOL is computed by counting the pixels belonging to its mask. To compute the lens instabilities, we propose to track the relative location of the lens segment inside the pupil. Hence, the pupil that is unstable due to the unconscious eye movements and surgical operations will be calibrated, and the relative position of the IOL is calculated by computing the distance between the centers of the IOL's mask and the pupil's mask, as shown in Figure \ref{fig:lens instability}.

\subsection{Lens Pose Estimation}

To compute the lens rotation amount, we propose estimating the lens orientation per frame based on the hooks' location. We use the Faster R-CNN \cite{Faster-R-CNN} framework for lens and hook localization. The Faster R-CNN network is a region-based CNN consisting of a backbone network and a region proposal network (RPN) followed by two branches: (1) a localization branch trying to output the most-fitted bounding box to each object and (2) a classification branch that detects the label associated with each detected object. We adopt several strategies to utilize the detection results in the inference stage. First, we only consider the detected bounding boxes with more than $60\%$ detection confidence. Considering that the IOL has only two hooks that are not always visible, we calculate the hooks' location using the detection results based on three scenarios: 

\begin{enumerate}
    \item If the network detects up to one hook fulfilling the determined threshold, the detected bounding box does not undergo any post-processing step.
    \item In case two hooks are detected by the network, we consider the position of the detected bounding boxes relative to the center of the IOL. If the angle between the two detected hooks is around 180 degrees, both detections are kept. Otherwise, only the detection with the higher confidence is considered a hook.
    \item In the condition that more than two bounding boxes fulfill the confidence threshold for the hook label, we perform hierarchical clustering with two clusters using the centers of the detected bounding boxes. Afterward, the bounding box with the highest detection confidence is selected as the best bounding box among all detections within each cluster. Having two bounding boxes, we further check their relative positions as described in scenario 2.
\end{enumerate}

To compute lens orientation, we calculate the angle of the line connecting the centers of the two hooks’ bounding boxes relative to the x-axis (in case of having two final hooks) or the angle of the line connecting the lens center and the visible hook’s bounding box relative to the x-axis (in case of having only one final hook). In the last case, we also consider the location of the detected hook relative to the lens center in calculating the lens orientation. 

\subsection{Lens Statistic Computation}
\label{sec: lens statistic computation}

\begin{algorithm}[t!]
    \caption{IOL Evaluation Pseudocode.}
    \label{algorithm}
    \SetKwInOut{Input}{Input}
    \SetKwInOut{Output}{Output}    
    \Input{The trained \textit{phase recognition} network\\ (CNN-RNN);
    
    The trained \textit{Semantic Segmentation} network\\ (AdaptNet);
    
    The trained \textit{Pose Estimation} network\\ (Faster R-CNN \cite{Faster-R-CNN}).
    }
    \Output{Lens correlation results.}
    \For{lens brand in $\{Technis, AvanSee, NC1, XC1\}$}{
    \For{video in lens group}{
       Feed the video to the phase recognition network to detect the implantation phase;
  
       Feed the frames of the post-implantation phase to the segmentation network to achieve the lens and pupil segmentation results;
   
       Refine the segmentation results by drawing a convex polygon around the detected masks;
   
       Compute lens-unfolding time based on the lens area (Eq. \eqref{eq: lens unfolding delay});
   
       Compute lens instability based on the relative position of the lens inside the pupil (Eq. \eqref{eq: lens instability});
   
       Feed the frames starting from the computed unfolding time until the end of surgery to the pose estimation network and compute lens orientation;
       
       Compute lens rotation based on lens orientation results (Eq. \eqref{eq: lens rotation});
       }
   Compute the correlation between lens unfolding delay and lens rotation (Eq. \eqref{eq: correlation});
   }
   \For{Pair of lens brands}{
   Perform three T-tests to evaluate the statistical differences between unfolding delay, instability, and rotation of the two groups (Eq. \eqref{eq: T-test}).
   }       
\end{algorithm}
The statistics of the IOL are calculated based on the lens' pose, visible area, and relative position. Supposing that we have the refined masks of the pupil ($\mathcal{M}^p = \{\mathcal{M}^p_1, ..., \mathcal{M}^p_n\}$) and IOL ($\mathcal{M}^l = \{\mathcal{M}^l_1, ..., \mathcal{M}^l_n\}$) starting from the post-implantation phase until the end of surgery with the rate of 25 \textit{fps} ($n = 25\times(t_{surgery}-t_{post-implantation})$), the masks' centers ($\mathcal{C}$) and areas ($\mathcal{A}$) can be denoted as:
\begin{equation}
Pupil 
\begin{cases}
    \mathcal{C}^p = [\mathcal{C}^p_1, \mathcal{C}^p_2, ..., \mathcal{C}^p_n] &\\
    \mathcal{A}^p = [\mathcal{A}^p_1, \mathcal{A}^p_2, ..., \mathcal{A}^p_n] &
\end{cases}
\end{equation}
\begin{equation}
IOL 
\begin{cases}
    \mathcal{C}^l = [\mathcal{C}^l_1, \mathcal{C}^l_2, ..., \mathcal{C}^l_n] &\\
    \mathcal{A}^l = [\mathcal{A}^l_1, \mathcal{A}^l_2, ..., \mathcal{A}^l_n] &
\end{cases}
\end{equation}

Where $\mathcal{A}_1$ and \textcolor{black}{$\mathcal{C}_1 = [\mathcal{C}_{1x}, \mathcal{C}_{1y}]$} correspond to the area and center of the masks in the first frame of the post-implantation phase, respectively. To mitigate the effect of the lens mask's prediction error on lens unfolding time prediction, we pass the lens area vector ($\mathcal{A}^l$) through a mean filter with a window size of 15 frames:
\begin{equation}
    \Tilde{\mathcal{A}}^l =
    \begin{cases}
        \frac{1}{15}\sum_{i-7}^{i+7}\mathcal{A}^l_i & 8<i<n-7\\
        \mathcal{A}^l_i & else
    \end{cases}
\end{equation}

For lens unfolding time ($t_U$), we compute the difference between the starting time of the post-implantation phase and when the frame-averaged visible lens' area is maximum for the first time:
\begin{equation}
    t_U = \argmax_t(\Tilde{\mathcal{A}}^l)
    \label{eq: lens unfolding delay}
\end{equation}

Lens instability ($Ins$) is computed based on the sum of the lens' absolute relative movements inside the pupil as follows:
\begin{equation}
    Ins = \sum_{i=1}^{n-1} ||\mathcal{C}^l_{i+1} - \mathcal{C}^p_{i+1}|-|\mathcal{C}^l_i - \mathcal{C}^p_i||.
    \label{eq: lens instability}
\end{equation}
Besides, considering the vector of lens orientations as $\mathcal{O}^l = [\mathcal{O}^l_1, ..., \mathcal{O}^l_n]$, we compute lens rotation based on the sum of absolute relative lens orientation changes, starting from the time when the lens is unfolded based on lens unfolding results:
\begin{equation}
    \mathcal{R}^l = \sum_{i=t_U}^{n-1}  |\mathcal{O}^l_{i+1} - \mathcal{O}^l_{i}|
    \label{eq: lens rotation}
\end{equation}

\subsection{Correlation Analysis}
\label{sec: statistical analysis}
We adopt the Pearson correlation coefficient to evaluate the correlations between lens unfolding delay and rotation. Assuming two subsets $x \subset \mathcal{X}$ and $y \subset \mathcal{Y}$ to be the representatives of unfolding delays and rotations of \textcolor{black}{a particular brand of} IOLs ($x = t_U$ and $y = R^l$), the Pearson correlation coefficient can be calculated as the covariance of $x$ and $y$ divided by the multiplication of standard deviation of these two sets:
\begin{equation}
\mathcal{P}earson_{x,y} = \frac{\sum_{i=1}^m (x_i-\overline{x})(y_i-\overline{y})}{\sqrt{\sum_{i=1}^m(x_i-\overline{x})^2\times\sum_{i=1}^m(y_i-\overline{y})^2}}
\label{eq: correlation}
\end{equation}

To evaluate the significant differences between the rotations of different IOL brands, we employ a t-test as follows:
\begin{equation}
    T = \frac{\overline{x} - \overline{y}}{\sqrt{\frac{1}{m}\times \sum_{i=1}^m(x_i-\overline{x})^2\times\sum_{i=1}^m(y_i-\overline{y})^2}}
    \label{eq: T-test}
\end{equation}

In the last two equations, $m$ is the number of samples in each set, being equal to \textcolor{black}{94} in our experiments.
\section{Experimental Setup}
\label{sec: experimental setup}

In this section, we first describe the datasets prepared and utilized for training the three neural network architectures as well as our large-scale IOL evaluations using the trained networks and the proposed framework in \ref{subsection: Video and Image Datasets}. Afterward, we explain the network training settings in \ref{subsection: Training Settings} and introduce the evaluation metrics in \ref{subsection: Evaluation Metrics}. The inference configurations are explained in \ref{subsection: Inference}.

\subsection{Video and Image Datasets}
\label{subsection: Video and Image Datasets}

This study adhered to the tenets of the Declaration of Helsinki with the approval of the ethics committee \textcolor{black}{(EK 28/17)}. All patients provided written informed consent before the study. The studies follow the reporting guidelines of the Standards for Quality Improvement Reporting Excellence (SQUIRE) and the Standards for Reporting of Diagnostic Accuracy (STARD). No patient received compensation or was offered any incentive for participating in this study.

The dataset used in this study stems from the Cataract-1k video dataset, containing 1000 videos of cataract surgeries recorded at the Klinikum Klagenfurt in 2020-2021.
\textcolor{black}{Since} the task of intraocular lens qualification involves three independent trained networks, namely \textcolor{black}{\textit{phase recognition}, \textit{pose estimation}, and \textit{semantic segmentation}, we} prepared three training/testing datasets. It should be noted that all training and testing images and videos are split patient-wise, meaning no images in the training and testing sets are sampled from the same video. This separation is intrinsic to building models working in real-world conditions. In addition to the training dataset, we use a large-scale dataset for validation, which we denote \textit{inference dataset}. The four mentioned datasets are explained as follows\footnote{\textcolor{black}{The annotated datasets will be publicly released in \href{https://ftp.itec.aau.at/datasets/ovid/lens-dislocation/}{https://ftp.itec.aau.at/datasets/ovid/lens-dislocation/.}}}:
\begin{enumerate}
    \item \textbf{Phase recognition dataset}: Contains annotations of the implantation phase versus the rest of the phases using the first and last implantation phase frames for 100 cataract surgery videos. 
    \item \textbf{Pose estimation dataset}: Includes bounding box annotations of the IOL and its hooks. Overall, 532 frames from 45 videos with the condition that at least one of the two lenses' hooks is visible were manually selected from the post-implantation phase. Next, 532 bounding boxes of the lenses and 821 bounding boxes of the lens' hooks were manually annotated. From these frames, 409 frames are used for training, and the remaining frames are used for testing.
    \item \textbf{Semantic segmentation dataset}: This dataset includes the segmentation masks of the IOL and pupil. The lens dataset contains 401 frames from 27 videos, and the pupil dataset contains 189 frames from 16 videos. From these annotations, we use 13 videos containing 141 frames with pupil annotation and 21 videos containing 292 frames with lens annotation for training. The remaining frames are utilized for testing.
    \item \textbf{Inference dataset}: Includes 376 other videos of cataract surgery and is used for validating trained models against different lenses' statistics. This dataset contains lenses from four brands: Technis, AvanSee, NC1, and XC1. From each brand, 94 videos were included. 
\end{enumerate}

\subsection{Training Settings}
\label{subsection: Training Settings}
For the phase recognition stage, all networks are trained for 20 epochs, with initial learning rates of 0.0002 for the VGG19 backbone and 0.0004 for the Resnet50 backbone. The learning rates are halved after ten epochs. 

For the semantic segmentation task, all networks are trained for 30 epochs. To account for differences in segmentation networks used for evaluations, three different initial learning rates ($lr_0\in{0.0005, 0.001, 0.002}$) are used, with a learning rate decrease of 0.8 every other epoch, and the results with the highest Dice coefficient are reported for each network. 

For the pose estimation task, we use the ResNet101 backbone and set the initial learning rate to 0.001. \textcolor{black}{This network is trained for 50 epochs.}

The backbones of all networks evaluated for phase recognition, lens/pupil semantic segmentation, and lens/hook localization are initialized with ImageNet~\cite{imageNet} weights. The input image size for all networks is set to $512\times 512\times 3$.
Data augmentation techniques, including motion blur, Gaussian blur, random contrast, random brightness, shift, scale, and rotation, are applied to prevent overfitting and improve generalization performance.
For the phase recognition and object localization tasks, we use binary cross entropy and the region-proposal-network (RPN)loss, respectively. For the semantic segmentation task, we adopt the cross-entropy-log-dice loss function consisting of categorical cross entropy and the logarithm of the soft Dice coefficient as follows:
\begin{equation}
\begin{split}
\mathcal{L} &= \lambda\times CE(\mathcal{X}_{Pred},\mathcal{X}_{True})\\
&- (1-\lambda)\times \log_2 Dice(\mathcal{X}_{Pred},\mathcal{X}_{True})
\end{split}
\label{eq:loss}
\end{equation}
Where \textit{CE} stands for \textit{Cross Entropy}, and $\mathcal{X}_{Pred}$ and $\mathcal{X}_{True}$ are the predicted masks and ground-truth segmentations, respectively. Besides, $\lambda$ is set to 0.8 in our experiments.

\subsection{Evaluation Metrics}
\label{subsection: Evaluation Metrics}
We evaluated the performance of each module separately. Phase recognition performance is evaluated using the standard classification metrics, namely Precision, Recall, F1-Score, and Accuracy. We use the Dice coefficient and intersection over union (IoU) to evaluate the semantic segmentation performance. For pose estimation evaluation, we adopt two schemes: (1) using the mean average precision (mAP) metric to measure the performance in object detection and localization, and (2) using the orientation detection error.

\subsection{Inference}
\label{subsection: Inference}
\textcolor{black}{Our proposed framework, leveraging the existing neural network architectures, is capable of real-time performance with the support of four RTX3090 GPUs. It's crucial to underscore the adaptability of our framework, recognizing the diversities in hardware infrastructures. In the case of lower hardware configurations, our algorithm allows for a seamless substitution of neural networks—such as those handling phase recognition, semantic segmentation, and object localization—with lightweight architectures. This ensures faster inference times without seriously compromising performance.}

\begin{table*}[t!]
\renewcommand{\arraystretch}{1.1}
\caption{Stepwise evaluation of the proposed lens irregularity detection framework.}
\label{tab:framework-evaluation}
\centering
\begin{tabular}{p{1.5cm}p{1.4cm}p{1.4cm}p{1.4cm}p{1.4cm}p{1.4cm}p{1.4cm}p{1.4cm}p{1.4cm}}
\specialrule{.12em}{.05em}{.05em}
\rowcolor{shadecolor}
\multicolumn{9}{l}{(A)	Phase recognition results of the end-to-end recurrent convolutional neural networks}\\\hline
 & \multicolumn{4}{c}{Backbone:VGG16} & \multicolumn{4}{c}{Backbone:ResNet50}\\\cmidrule(lr){2-5}\cmidrule(lr){6-9}
RNN & Precision & Recall & F1-Score & Accuracy & Precision & Recall & F1-Score & Accuracy\\
\specialrule{.12em}{.05em}{.05em}
GRU \cite{GRU}& $0.97$ & $0.96$ & $0.96$ & $0.96$ & $0.90$ & $0.94$ & $0.94$ & $0.94$\\
LSTM \cite{LSTM}& $0.98$ & $0.98$	& $0.98$ & $0.98$ & $0.96$	& $0.96$ & $0.96$ &	$0.96$\\
BiGRU & $0.97$ & $0.96$	& $0.96$ & $0.96$ & $0.95$ & $0.95$ & $0.95$ &	$0.95$\\
BiLSTM & $\mathbf{1.00}$ & $\mathbf{1.00}$ & $\mathbf{1.00}$ & $\mathbf{1.00}$	& $0.98$ & $0.98$ & $0.98$ & $0.98$\\
\specialrule{.12em}{.05em}{.05em}
\end{tabular}

\begin{tabular}{p{1.5cm}p{1.4cm}p{1.4cm}p{1.4cm}p{1.4cm}p{1.4cm}p{1.4cm}p{1.4cm}p{1.4cm}}
\specialrule{.12em}{.05em}{.05em}
\rowcolor{shadecolor}
\multicolumn{9}{l}{(B)	Lens and pupil segmentation results based on intersection over union (IoU) and Dice}\\\hline
Object & Metric & U-Net \cite{U-Net}& PSPNet \cite{PSPNet} & CE-Net \cite{CE-Net} & CPFNet \cite{CPFNet}& UNet++\slash DS \cite{UNet++}& UNet++ \cite{UNet++}& AdaptNet \cite{LensID}\\\hline
\multirow{2}{*}{Lens} & IoU	& $61.89$ & $71.40$ & $70.56$ & $75.38$ &	$82.32$ & $83.61$ & $\mathbf{87.09}$\\
& Dice & $73.86$ & $81.53$	& $82.22$	& $85.26$	& $89.95$ & $90.44$ &	$\mathbf{92.62}$\\\hline
\multirow{2}{*}{Pupil} & IoU & $83.51$ & $89.55$ & $87.66$ & $92.33$ & $95.28$ & $96.02$ & $\mathbf{96.06}$\\
& Dice & $89.36$ & $94.18$	& $93.32$	& $95.99$	& $97.53$	& $97.96$	& $\mathbf{97.98}$\\
\specialrule{.12em}{.05em}{.05em}
\end{tabular}

\begin{tabular}{p{4.35cm}p{4.35cm}p{4.35cm}}
\specialrule{.12em}{.05em}{.05em}
\rowcolor{shadecolor}
\multicolumn{3}{l}{(C) Lens and hook localization results of Faster R-CNN \cite{Faster-R-CNN} }\\\hline
Backbone & mAP	& mAP@0.5IoU\\\hline
ResNet50 & $0.547$ & $0.828$\\
\specialrule{.12em}{.05em}{.05em}
\end{tabular}

\begin{tabular}{p{2.45cm}p{2.45cm}p{2.45cm}p{2.45cm}p{2.4cm}}
\specialrule{.12em}{.05em}{.05em}
\rowcolor{shadecolor}
\multicolumn{5}{l}{(D)	Lens orientation computation results}\\\hline
Mean Error	& Std of Error	& Top $75\%$ Error &	Top $50\%$ Error &	Top $25\%$ Error\\\hline
$3.707$ &	$7.499$ & $3.455$ & $1.31$ & $0.439$\\
\specialrule{.12em}{.05em}{.05em}
\end{tabular}

\end{table*}

\section{Experimentsal Results}
\label{sec: experimental results}
In this section, we first assess the effectiveness of each distinct module within the proposed framework based on the results reported in Table \ref{tab:framework-evaluation}. Next, we use the proposed framework with trained networks for the statistical evaluations of the four mentioned IOL brands to assess the possibility of automating IOL evaluation and post-operative IOL irregularity prediction.

\subsection{Surgical Phase Classification Results}
Table \ref{tab:framework-evaluation}-A presents the implantation phase recognition results of the utilized recurrent CNN considering two different backbone networks and four different recurrent layers. It can be perceived from the table that the utilized architecture can effectively capture the joint spatio-temporal features associated with the implantation phase disregarding the backbone network's model and the recurrent layer. Surprisingly, the network with the bidirectional LSTM layer and the VGG19 backbone could retrieve $100\%$ of the three-second clips belonging to the implantation phase. Moreover, this network was $100\%$ precise in discriminating the implantation phase versus the rest of the phases. These results confirm the effectiveness of the proposed approach in detecting the starting point for lens statistics computation.

\subsection{Semantic Segmentation Results}
To highlight the superiority of the utilized semantic segmentation network in segmenting the artificial lens and pupil, we have compared its results with several state-of-the-art networks. Table \ref{tab:framework-evaluation}-B lists the semantic segmentation results of the proposed approach (AdaptNet) and rival approaches based on the mean and standard deviation of the IoU and Dice coefficient. According to the IoU results, AdaptNet has achieved the best performance in segmenting both IOL and pupil. It gains at least $4\%$ relative improvement in IoU and $2.4\%$ relative improvement in Dice coefficient compared to the best alternative network (UNet++) in lens segmentation.

\subsection{Lens' Pose Estimation Results}
Table \ref{tab:framework-evaluation}-C lists the lens and hooks localization results based on mean average precision (mAP) and mean average precision at $50\%$ intersection over union ($mAP@0.5IoU$). The results indicate that the network is $82\%$ accurate in localizing the bounding boxes with at least $50\%$ intersection over union. 

According to Table \ref{tab:framework-evaluation}-D, the proposed lens orientation computation method shows a mean error equal to 3.707 degrees, with its standard deviation being equivalent to 7.499 degrees. Moreover, the model shows less than 1.5 degrees error in orientation computation for at least $50\%$ of samples in the test set. These results confirm the reliability of the proposed method for lens orientation computation.

\begin{table*}[t!]
\renewcommand{\arraystretch}{1.1}
\caption{Statistical analysis of the behavior of different intraocular lens brands. The statistically significant results are bold.}
\label{tab:lens-evaluation}
\centering
\begin{tabular}{p{3cm}p{2.5cm}p{2.5cm}p{2.5cm}p{2.5cm}}
\specialrule{.12em}{.05em}{.05em}
\rowcolor{shadecolor}
\multicolumn{5}{l}{(A)	Pearson correlation and p-values between lens unfolding delay and lens rotation}\\\hline
Lens Brand	& Tecnis & AvanSee & NC1 & XC1\\\hline
Pearson Correlation	& $0.0238$ & $0.2392$ & $0.3592$ & $0.2639$\\
P-Value	& $0.8194$ & $\mathbf{0.0202}$ & $\mathbf{0.0003}$	& $\mathbf{0.0101}$\\
\specialrule{.12em}{.05em}{.05em}
\end{tabular}

\begin{tabular}{p{3cm}p{2.5cm}p{2.5cm}p{2.5cm}p{2.5cm}}
\specialrule{.12em}{.05em}{.05em}
\rowcolor{shadecolor}
\multicolumn{5}{l}{(B)	P-values resulting from the t-test between the rotations of different IOLs}\\\hline
& Tecnis & AvanSee	& NC1 & XC1\\\hline
Tecnis	& N/A	& $\mathbf{0.0322}$ & $\mathbf{5.01 \times 10^{-6}}$ &	$\mathbf{8.26 \times 10^{-18}}$\\
AvanSee	& $\mathbf{0.0322}$ & N\slash A	& $\mathbf{0.0019}$ & $\mathbf {1.48 \times 10^{-14}}$\\
NC1	& $\mathbf{5.01 \times 10^{-6}}$ & $0.0019$	& N \slash A & $0.0363$\\
XC1	& $\mathbf{8.26 \times 10^{-18}}$ &	$\mathbf{1.48 \times 10^{-14}}$ &	$0.0363$	& N \slash A\\
\specialrule{.12em}{.05em}{.05em}
\end{tabular}

\begin{tabular}{p{3cm}p{2.5cm}p{2.5cm}p{2.5cm}p{2.5cm}}
\specialrule{.12em}{.05em}{.05em}
\rowcolor{shadecolor}
\multicolumn{5}{l}{(C)	P-values resulting from the t-test between the unfolding of different IOLs}\\\hline
& Tecnis & AvanSee	& NC1 & XC1\\\hline
Tecnis	& N\slash A & $0.143$ & $0.798$ & $0.054$\\
AvanSee	& $0.143$	& N \slash A & $0.204$ & $0.471$\\
NC1	& $0.798$	& $0.204$	& N \slash A & $0.116$\\
XC1	& $0.054$	& $0.471$	& $0.116$	& N\slash A\\
\specialrule{.12em}{.05em}{.05em}
\end{tabular}

\end{table*}

\subsection{IOL Evaluations and Statistical Comparisons}
\begin{figure*}[!tb]
    \centering
    \includegraphics[width=1\textwidth]{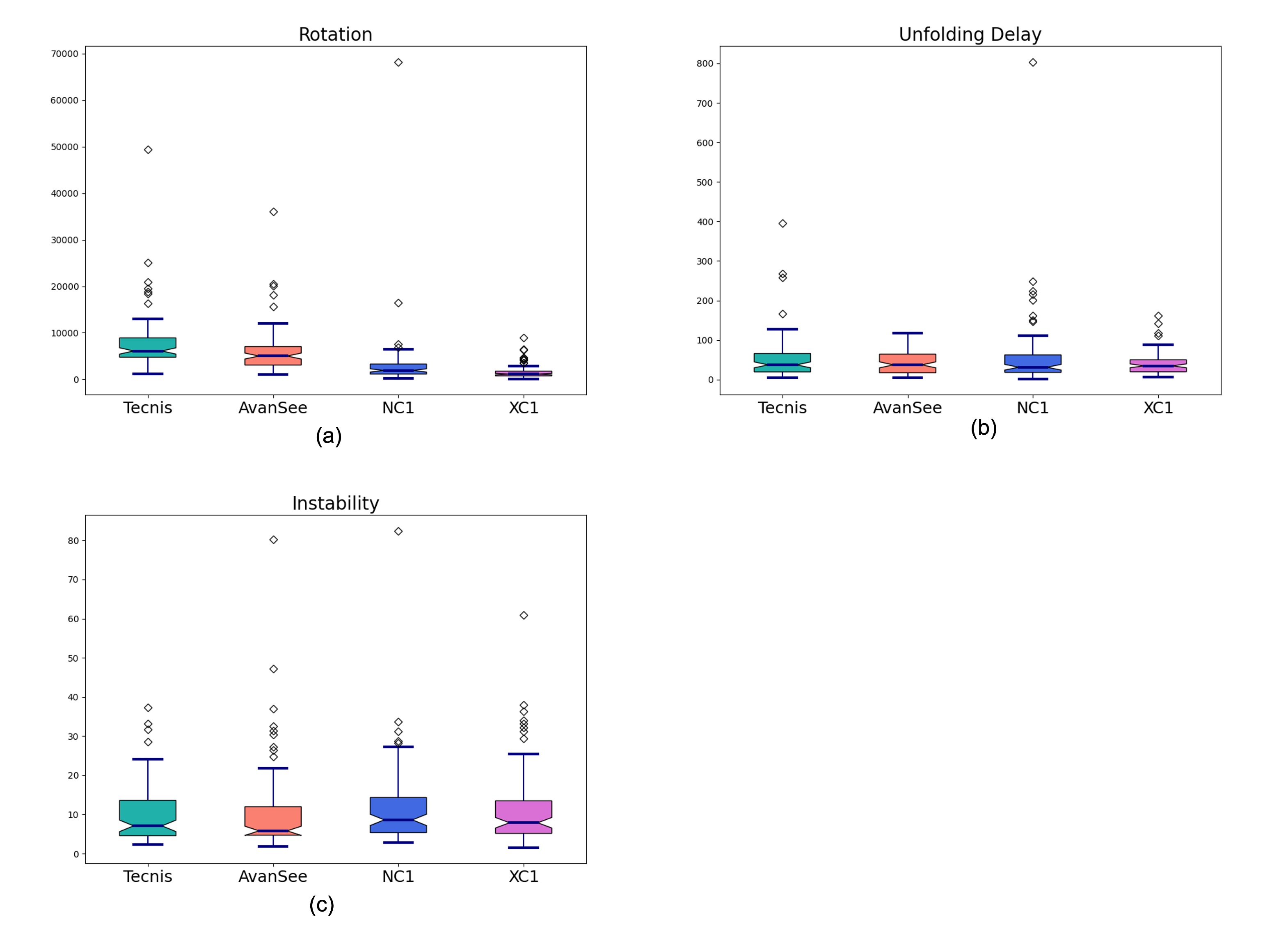}
    \caption{The lens statistics for one representative cataract surgery video.}
    \label{fig:BoxPlot}
\end{figure*}

In order to statistically compare the behavior of different groups of lenses, we have computed lens unfolding time, instability, and unfolded-lens rotation (rotation for short) for \textcolor{black}{brands} of intraocular lenses. Figure \ref{fig:BoxPlot} demonstrates the boxplots of the three mentioned measurements for each group of lenses containing 94 cataract surgery videos. 
Regarding intra-operative rotation after unfolding, the four brands of lenses show significantly different behaviors based on the middle quartile and the interquartile range and the amount of skewness. According to the unfolding plots, XC1 shows the smallest interquartile range (IQR) and the smallest overall spread ($[Q_1-1.5 \times IQR,Q_3+1.5 \times IQR]$), suggesting that XC1 lenses have less dispersed unfolding time. On the other hand, the Tecnis and NC1 boxplots have a substantially higher upper whisker and positive skew, meaning more dispersed unfolding time and non-normally distributed data. The distant outliers in the two latter lenses can also imply more irregularities. The boxplots of instability for all four groups of lenses have a relatively close length of interquartile range. We can conclude that intra-operative instability of the lens during unfolding cannot be used as an indicator of post-operative rotation. 

We infer from the boxplots that there is a higher statistical difference between the unfolding delay and rotation of the four types of IOLs compared to their instabilities. Hence, we have further computed the Pearson correlations between the unfolding delay and rotation of these lenses. As listed in Table \ref{tab:lens-evaluation}-A, the p-values for correlations between lens unfolding time and rotation of the AvanSee (0.0202), NC1(0.0003), and XC1(0.0101) are less than 0.05. This suggests that there is a significant correlation between the unfolding time and rotation of each mentioned group of IOLs. Besides, we have computed the p-value based on a t-test between the rotation of the four groups of IOLs (Table \ref{tab:lens-evaluation}-B). The t-test results confirm the statistically significant differences between the rotations of Tecnis vs. Avansee (0.0322), Tecnis vs. NC1 ($5.01\times 10^{-6}$), Tecnis vs. XC1 ($8.26\times10^{-18}$), AvanSee vs. NC1 (0.0019), AvanSee vs. XC1 ($1.48\times10^{-14}$), and NC1 vs. XC1 (0.0363). The results of the t-test between the unfolding time of different IOLs in Table \ref{tab:lens-evaluation}-C also indicate a nearly significant difference between the unfolding delay of the XC1 and Tecnis lenses (0.054).

\section{Discussion}
\label{sec: discussion}

\textcolor{black}{We formulated a hypothesis suggesting that variations in the unfolding delay of intraocular lenses (IOLs) play a significant role in their susceptibility to post-operative rotational instability. To substantiate this idea, we conducted a comparative analysis of the unfolding time in our studies and cross-referenced it with data published on post-operative IOL rotation, as indicated by references \cite{SCHARTMULLER202072,RotStab,Schartmuller186}.}

\textcolor{black}{In the study that compared the post-operative rotational stability of the Hoya Vivinex lens (XC1) and the Tecnis IOL, conducted by Osawa et al. \cite{RotStab}, it was revealed that the Hoya Vivinex lens displayed less early post-operative rotation than the Tecnis IOL. Our research identified distinct differences in the unfolding behavior of these two lenses, with the Tecnis IOL exhibiting a less predictable unfolding time and more irregularities when compared to XC1. Furthermore, a recent cohort study involving 647 implanted Tecnis IOLs highlighted a noteworthy post-operative rotation issue, with $8.1\%$ of cases experiencing rotations exceeding 5 degrees and $3.1\%$ requiring secondary interventions for re-positioning \cite{SCHARTMULLER202072}. This pattern of high post-operative absolute rotation in Tecnis lenses has also been corroborated by other studies, as cited in reference \cite{Schartmuller186}.}

\textcolor{black}{Our proposed framework has effectively generated features that confirm statistically significant differences in the intra-operative rotational stability of various IOL brands. The direct correlation between the intraoperative rotation of a pair of lenses (Tecnis vs. XC1) and their subsequent post-operative rotational stability reaffirms the predictability of this complication during surgery. These findings can provide valuable guidance to surgeons in their IOL selection, particularly in cases where patient-specific factors, such as myopia (where lens dislocation is more likely and lens rotation can significantly impact vision, as discussed in \cite{lee2020risk}), need to be considered. Ultimately, our research contributes to the prevention of post-operative IOL complications, enhancing patient outcomes and satisfaction.}

\textcolor{black}{\subsection{Limitations of the Present Study}
Evaluating the intraoperative behavior of IOLs can empower the predictability of the post-operative IOL’s dislocation. However, in addition to the lens characteristics, intrinsic factors such as physicochemical and surface properties of acrylic IOL, modifiable factors such as the temperature of the IOL at the time of implantation, and the viscoelastic used can play a role in the unfolding of an IOL during cataract surgery. The proposed framework enables surgeons to evaluate the intraoperative behavior of IOLs and correlate them with IOLs’ early postoperative stability. These results help the surgeons measure each factor’s influence on IOL post-operative complications.}

\section{Conclusion}
\label{sec: conclusion}

\textcolor{black}{Intraocular lens dislocation stands as a pivotal postoperative complication in cataract surgery, warranting considerable attention due to its implications on patient outcomes and healthcare costs. This paper presents a pioneering framework, representing the first endeavor to automate the extraction of critical lens statistics including intraocular lens unfolding delay, instability, and rotation during cataract surgery. We have proposed, evaluated, and utilized a CNN-RNN-based framework for a large-scale evaluation of four brands of IOLs. These results have enabled us to measure statistical correlations between different features in each IOL and differences in the behavior of four commonly used IOL brands during the surgery. The proposed framework not only helps enhance our understanding of IOL behavior during cataract surgery but also offers a major step toward predicting and ultimately preventing such a crucial irregularity. By improving the predictability of complications such as lens dislocation, we can substantially reduce the economic burden for patients and healthcare systems alike. Beyond the financial aspect, preventing such complications translates to heightened patient satisfaction, improved quality of care, and overall enhanced surgical experiences.}

\textcolor{black}{Expanding upon our current work, future investigations should focus on an exhaustive examination of the multifaceted factors influencing the intraoperative behavior of intraocular lenses (IOLs) during cataract surgery. This could encompass a comprehensive study of the relationships between lens characteristics, the inherent properties of acrylic IOLs, the temperature during implantation, and the type of viscoelastic used. 
In addition to this multifaceted exploration, the practical applicability of our framework for real-time analysis of IOL behavior during surgical procedures is reliant on the existing hardware infrastructure within operating rooms.
Accordingly, a key consideration for enhancing the usability of such frameworks is the reduction of dependency on powerful GPUs by optimizing network parameters. This optimization will not only improve the efficiency of our system but also promote its wider adoption across diverse surgical environments.}

\section*{Acknowledgments}
The investigators were independent of the funders. Negin Ghamsarian and Klaus Schoeffmann have full access to the data and can take responsibility for the integrity of the data and accuracy of the data analysis. The lead author affirms that the manuscript is an honest, accurate, and transparent account of the study being reported; that no important aspects of the study have been omitted; and that any discrepancies from the study as planned have been explained.

Data sharing: all datasets prepared and used in this study can be shared for further scientific investigations upon request.

\bibliographystyle{acm}
\bibliography{bibtex.bib}

\end{document}